# Path2Models: Large-scale generation of computational models from biochemical pathway maps

Finja Büchel[1,2,*], Nicolas Rodriguez[1,3,*], Neil Swainston[4,*], Clemens Wrzodek[2,*], Tobias Czauderna[5], Roland Keller[2], Florian Mittag[1,2], Michael Schubert[1], Mihai Glont[1], Martin Golebiewski[6], Martijn van Iersel[1], Sarah Keating[1], Matthias Rall[2], Michael Wybrow[7], Henning Hermjakob[1], Michael Hucka[8], Douglas B. Kell[4,9], Wolfgang Müller[6], Pedro Mendes[4,10,11], Andreas Zell[2], Claudine Chaouiya[12], Julio Saez-Rodriguez[1], Falk Schreiber[5,13], Camille Laibe[1], Andreas Dräger[2,14], Nicolas Le Novère[1,3,§]

*Affiliations*:

[1]European Bioinformatics Institute, Wellcome Trust Genome Campus, Hinxton, Cambridge, United Kingdom

[2]Center for Bioinformatics Tuebingen (ZBIT), University of Tuebingen, Tübingen 72076, Germany

[3]Babraham Institute, Babraham Research Campus Cambridge, United Kingdom

[4]Manchester Institute of Biotechnology, The University of Manchester, M1 7DN, United Kingdom

[5]Leibniz Institute of Plant Genetics and Crop Plant Research, Gatersleben D-06466, Germany

[6]HITS gGmbH, D-69118 Heidelberg, Germany

[7]Caulfield School of Information Technology, Monash University, Victoria 3800, Australia

[8]Department of Computing and Mathematical Sciences, California Institute of Technology, Pasadena, CA 91125 USA

[9]School of Chemistry, The University of Manchester, Manchester M13 9PL, United-Kingdom

[10]School of Computer Science, The University of Manchester, Manchester M13 9PL, United-Kingdom.

[11]Virginia Bioinformatics Institute, Virginia Tech, Blacksburg, Virginia, USA

[12]Instituto Gulbenkian de Ciência (IGC), Oeiras P-2780-156, Portugal

[13]Institute of Computer Science, University of Halle-Wittenberg, Halle, Germany

[14]Present address: University of California, San Diego, Bioengineering Department, La Jolla, CA 92093-0412, USA

* These authors contributed equally to the work and are listed alphabetically.

[§]To whom correspondence should be addressed. lenov@babraham.ac.uk

## Abstract

**Background**: Systems biology projects and omics technologies have led to a growing number of biochemical pathway models and reconstructions. However, the majority of these models are still created *de novo*, based on literature mining and the manual processing of pathway data.

**Results**: To increase the efficiency of model creation, the Path2Models project has automatically generated mathematical models from pathway representations using a suite of freely available software. Data sources include KEGG, BioCarta, MetaCyc and SABIO-RK. Depending on the source data, three types of models are provided: kinetic, logical and constraint-based. Models from over 2 600 organisms are encoded consistently in SBML, and are made freely available through BioModels Database at http://www.ebi.ac.uk/biomodels-main/path2models. Each model contains the list of participants, their interactions, the relevant mathematical constructs, and initial parameter values. Most models are also available as easy-to-understand graphical SBGN maps.

**Conclusions**: To date, the project has resulted in more than 140 000 freely available models. Such a resource can tremendously accelerate the development of mathematical models by providing initial starting models for simulation and analysis, which can be subsequently curated and further parameterized.

# BACKGROUND

Since the discovery of the set of biochemical transformations known as the Embden-Meyerhof-Parnas glycolysis pathway in the early twentieth century, the concepts of pathways and networks have become useful and ubiquitous tools in the understanding of biochemical processes. Biochemical pathways provide a qualitative representation of chains of molecular interactions and chemical reactions that are known to take place in cells. Such interactions result in changes in the concentration, state or location of chemical entities. Pathways aim at providing a detailed representation of this biochemical reality, based on observations of the reactions. As such, the elucidation of biochemical pathways is being dramatically sped up with the efforts of molecular biology and biochemistry research, and particularly with the recent appearance of high-throughput omics technologies

The definition of biochemical pathways is largely arbitrary, as in practice they are interlinked and interdependent in the functioning cell. Nevertheless, it is convenient to partition these pathways into different types such as signaling pathways, metabolic networks, gene regulatory networks, etc. With the growing number and complexity of biochemical pathways, a number of public databases have attempted to catalog them and provide access to their computational representation. These well-curated resources include MetaCyc [1], KEGG [2], the Nature Pathway Interaction Database (PID) [3], Reactome [4] and WikiPathways [5].

While such resources remain extremely useful, they provide purely qualitative, static, representations of molecular interactions. Although such representations can be used in the context of experimental data mapping and interpretation [6], they fail to provide a quantitative understanding of cellular mechanisms. A key to the understanding of biological processes is to go beyond mere accumulation of observations, even on the large scale as in multi-omics data collection, and to move towards their quantitative prediction. This understanding can in turn lead to the alteration of biological processes, for instance through pharmaceutical intervention, and even to the design of entirely novel processes in the fields of metabolic engineering and synthetic biology. Accordingly, over the last decade and a half, the increased availability of quantitative experimental data has motivated scientists to develop predictive and quantitative representations of pathways and entire networks in the form of computational models.

Computational models rely on mathematical frameworks to describe the structures and behaviors of systems. A model consists of variables, functions and constraints. Different types of models exist, such as kinetic models, logical models, rule-based models, multi-agent models, statistical models and many more. In contrast to most pathways, which seek to provide detailed representations of biochemical knowledge, models can be more abstract representations of the reality, depending on the needs of the modeler, the experimental data available and the investigation being undertaken. Models can therefore exhibit different levels of granularity for the variables and different degrees of precision for the mathematical functions. Computational models of biochemical systems are shared through databases such as BioModels Database [7] and the CellML repository [8], with their storage and exchange relying heavily on the adoption of standard formats such as the Systems Biology Markup Language (SBML [9]) and the Systems Biology Graphical Notation (SBGN [10]).

Different types of models can be generated from pathway databases. Biochemistry, and in particular metabolism, is very often represented using *process descriptions*. *Processes* are the biochemical reactions and transport processes between compartments that transform nominally homogeneous pools of biochemical entities into other pools of entities. In process descriptions, a pathway is a bipartite graph formed of the biochemical entities and the processes that consume or produce them. Models based on process descriptions can be encoded with the elements of SBML *Core* and represented in the *Process Description* language of SBGN [10].

Quantitative methods for modeling biological networks require accurate knowledge of the biochemical reactions, their stoichiometric and kinetic parameters, and in the case of metabolic pathway modeling [11], initial concentrations of metabolites [12] and enzymes [13]. In many cases, such experimentally derived parameters are unavailable. This has led to the development of several qualitative approaches, based on influence networks rather than process descriptions. Examples are logical modeling in multiple variants, from Boolean or multi-valued networks [14-16] to discrete algebra [17] and differential equations [18], Petri nets [19] and predicate logic [20]. Qualitative models typically refer to regulatory or signaling networks, and are based on the definition of an influence or signal-flow graph, rather than the depiction of consumption and production of pools of entities. These methods have proven useful in recent years in the interpretation of data from perturbation experiments, phosphoproteomics and gene expression studies [21]. SBML has recently been extended to support such logical models, which can be encoded with the newly introduced *Qualitative Models* package for SBML Level 3 (henceforth abbreviated as the SBML *qual* package [22]) and represented in the *Activity Flow* language of SBGN.

In addition to curated pathway databases, the availability of well-annotated entire genomes, together with methods for reconstructing and constraining large-scale biochemical networks, has led to the reconstruction of comprehensive metabolic pathways, including all enzymes known to be encoded by an organism. The development of these genome-scale metabolic network reconstructions, and their analysis through constraint-based modeling approaches, is becoming increasingly widespread in driving the understanding of metabolism in a diverse range of organisms. The number of such genome-scale metabolic reconstructions published over the last ten years has grown considerably, with over 50 such reconstructions recently reported [23], covering a range of single- and multicellular organisms.

Metabolic reconstructions attempt to provide a computational and mathematical representation of the metabolic capabilities of the cell. Reconstructions have been used in a number of research topics including metabolic engineering, genome-annotation, evolutionary studies, network property analysis, and interpretation of omics datasets [24]. The development of genome-scale metabolic reconstructions typically involves a labor-intensive, manual process, with timescales of up to two years reported for their production [25]. While it is recognized that the development of high-quality metabolic reconstructions requires significant curation, and is dependent upon manual [26-30] or semi-automated literature mining [31,32], there have been notable recent steps towards semi-automation of the reconstruction process, which aim to reduce the number of tasks that must be performed manually.

Traditionally, computational models have been painstakingly (and manually) built from primary information obtained from the literature and from dedicated experiments. Because of the increasing size and complexity of these models, this approach is no longer sustainable. Modelers have therefore begun to build models directly based on data imported from pathway databases. However, until recently, this has mostly been done on a tedious case-by-case basis and repeated separately by different researchers because the results were not shared in a consistent fashion. The Path2Models project attempts to mitigate this often duplicated initial modeling step by generating computational models from pathways on a large scale, applying consistent, community-developed and well-supported data formats, and to make the results available to the community as a whole.

This manuscript therefore describes the conversion of pathway information to computational models in a consistent and high-throughput manner. The Path2Models project has generated three types of models: quantitative, kinetic models of metabolic pathways; qualitative, logical models of non-metabolic (primarily signaling) pathways; and genome-scale metabolic reconstructions. The models are generated in SBML, and in many cases are augmented with visual representations in the form of SBGN documents. All of the models share a consistent format and are semantically annotated according to the Minimum Information Required In the Annotation of Models (MIRIAM) specification [33]. In practice, this means that all components of the models (metabolites, genes, enzymes, reactions, etc.) are tagged with unambiguous identifiers from publicly available, third party databases. The models can therefore be easily queried, compared, merged and expanded, and are immediately amenable to integration with experimental data [34]. The resulting models are made publicly available through BioModels Database [7] and can be used as starting point for further development.

# RESULTS

## Workflow from biochemical pathways to computational models

In order to generate computational models from biological pathways on a large scale, a software pipeline composed of several steps that can be run sequentially or in parallel was developed (**Figure 1**). The pathways must first be converted from their original format to a standard computer-readable format, which will be used throughout all subsequent steps of the pipeline. This work describes the conversion of pathway information from KEGG, MetaCyc, and BioPAX [35] into SBML models, lacking both mathematics and numerical values. These preliminary networks were then processed to annotate, merge, extend and complete them with mathematical expressions where possible. All software modules utilized in this work are freely distributed, and readers can re-use them on their own or within their own workflows.

Three parallel pipelines of data processing were implemented: 1) kinetic metabolic models represented by processes were encoded in SBML Level 3 *Core* format, enriched with modular rate-laws and depicted using SBGN *Process Descriptions;* 2) qualitative metabolic and non-metabolic (mostly signaling) pathways, represented as influence diagrams, were encoded in SBML using the Level 3 *qual* package, in a form ready for logical modeling and depicted using SBGN *Activity Flows*; 3) *g*enome-scale metabolism reconstructions were similarly encoded in SBML, in a format amenable to constraint-based modeling.

## Generation of quantitative kinetic process models from metabolic pathways

The metabolic pathways distributed by KEGG are described in terms of processes, and formed the basis of the process-based reconstructions. 112 898 maps describing up to 154 metabolic pathways in 1 514 organisms were converted into process description models encoded in SBML Level 3 *Core.* The resulting SBML documents were converted into SBGN *Process Descriptions* (PD) maps, in order to provide defined graphical representations of all models (**Figure 2**).

Reconstructions of metabolic networks were completed by the addition of experimentally determined rate laws and parameter values from the SABIO-RK database [36]. SABIO-RK is a reaction-kinetics database that contains experimentally obtained rate laws for a large collection of (bio-) chemical reactions, including measured parameter values and experimental conditions, such as the pH value or the temperature, under which the rate was measured [37]. It was therefore desirable to extract as much information from SABIO-RK as possible and relevant. For all reactions that lacked corresponding entries in SABIO-RK, the kinetic rate laws were inferred *ab initio* (see Methods). At the moment, the SABIO-RK database mainly focuses on a selection of relevant model organisms, for which many rate laws can already be extracted (see **Figure 3**), for instance, 12% for *Homo sapiens*, 10% for *Rattus norvegicus*, and 8% for *Escherichia coli*. Across the full range of organisms we considered, 6204 reactions (0.22%) could be equipped with rate laws from SABIO-RK.

## Generation of qualitative models from signaling pathways

From the KEGG pathway database, 27 306 maps describing 167 non-metabolic pathways in 1 514 organisms were converted into influence maps models encoded with the SBML Level 3 *qual* package.

Prior to our use to convert non-metabolic pathways, no attempt had been made to encode pathway models using the SBML *qual* syntax. We uncovered several aspects of the package specification that caused problems when applied to actual pathways and the project provided a valuable concrete situation to help resolve these issues. For example, the information available originally permitted the description of interaction graphs but was not sufficient to define logical rules specifying the effects of combined interactions. This led to the introduction of a sign attribute for indicating whether a given interaction has a positive, negative or unknown effect. This can then be used as a constraint to parameterize a logical model further. The project therefore accelerated the development and finalization of the SBML Level 3 *qual* specification,

KEGG relations sometimes consist exclusively of the subtypes phosphorylation, dephosphorylation, glycosylation, ubiquitination, or methylation. These relations cannot be interpreted in terms of positive or negative influences on a transition (for instance, a phosphorylation can increase or decrease the activity of a protein). In those cases, the *sign* attribute was initially set to *unknown* for the *input* element of the corresponding *transition*. Whenever possible, the KEGG pathways were augmented with interaction information imported from the BioCarta pathways distributed by the Nature Pathway Interaction Database (PID) [3]. PID provides human pathways in the BioPAX format Level 3, which specifies a *ControlType* attribute for each interaction. The *ControlType* attribute determines whether the interaction represents activation or inhibition. With the additional information from the PID, it was possible to extend 35 human pathways.

**Genome-scale metabolic reconstructions**

Genome-scale metabolic reconstructions of 2 630 organisms were generated through extraction of pathway data from the KEGG and MetaCyc databases using an updated version of the pre-existing software libAnnotationSBML and the SuBliMinaL Toolbox [38,39]. All reconstructions contain data from KEGG, and many of these have been augmented with data from MetaCyc for the corresponding organism. In each case, MNXref was used to reconcile metabolite and reaction identifiers across the different data resources [40]. As well as providing mapping of KEGG and MetaCyc identifiers, MNXref also applies a default metabolite formula and charge state according to an assumed pH of 7.3, and ensures mass and charge balancing of reactions where possible. Furthermore, MNXref provides mapping to additional identifiers, which have been extracted and incorporated into the collection of genome-scale reconstructions. As such, as well as ensuring consistent metabolite and reaction identifiers across all 2 630 reconstructions, all models also contain identifier cross references to numerous commonly used resources, including BiGG [41] and the Model SEED [42], further enhancing their interoperability.

A minimal growth medium (consisting of a single carbon source, glucose), appropriate transport reactions, and 30 common biomass components were specified in each model, including all 20 amino acids, RNA and DNA nucleotide precursors, glycogen and ATP (see Methods). A default biomass objective function was added, containing these components, with the intention of facilitating subsequent analysis and curation. The models were then formatted such that they could be analyzed with a range of SBML-compatible software tools, including the COBRA Toolbox [43,44]. **Figure 4** describes the workflow that was used in the automated reconstruction process.

The resulting 2 630 models range in size from the smallest, *Candidatus Tremblaya princeps PCVAL*, containing 131 metabolites and 63 metabolic reactions, to *Homo sapiens*, with 3 270 metabolites and 3 416 metabolic reactions. All models were analyzed for their ability to synthesize each defined biomass precursor from the minimum growth medium, taking into account reaction directionalities specified in KEGG and/or MetaCyc where available. Of these, only the model of *Drosophila melanogaster* was able to synthesize all specified 30 biomass components. The *Homo sapiens* model was incapable of synthesizing the amino acids cysteine, histidine, isoleucine, leucine, lysine, methionine, threonine, tryptophan and valine. Of these, all but cysteine are known essential amino acids. Additionally, the model is unexpectedly able to synthesize phenylalanine, an essential amino acid. Nevertheless, these analysis results indicate that the draft model is largely predictive of the amino acid essentiality, with the anomalies of cysteine and phenylalanine synthesis pathways providing starting points for manual curation.

The full results of this study are provided in a definitive list of all models produced is in **Supplementary Table 1**. The results can also be viewed as a phylogenetic tree, generated by the Integrated Tree Of Life (iTOL) web application [45], at [46] (see **Figures 5** and **6**).

**Access to the resulting knowledge base**

BioModels Database is the reference repository of computational models of biological interest encoded in SBML. This resource allows biologists to store, search, retrieve and display mathematical models. One of the main qualities of the repository lies in its contents: all are distributed in standard formats and using a free license, allowing easy re-use. The models generated by the project have been made publicly available from BioModels Database since release 22 under the name "Path2Models" [47]. The size of the distribution of all these models is presented in **Figure 7**. A new branch in the model-processing pipeline was created in order to accommodate those models, as they are not expected to go through the usual manual curation and annotation phases. A dedicated search infrastructure for the Path2Models branch was provided with release 23. **Figure 8** presents the relative populations of the different topics, as compiled from the Gene Ontology annotation of the models. The Path2Models branch of BioModels Database is not considered to be a frozen resource, and improved versions will be released as they are made available.

# DISCUSSION

## Automatically generated models are only a starting point

The workflow described here enables the automatic generation of a large number of computational models from existing pathway data resources. The procedure is essentially the same as for building an individual model from the same data. However, instead of independent scientists enacting this procedure again and again as the needs arise, the initial data processing is performed in bulk. Scientists can then focus on the more interesting tasks of adapting the models to their questions, adding initial conditions and parameter values, and running simulations to answer biological questions in the organisms and/or pathways in which they are interested.

The added value provided by the initial models to such research activities largely depends on the quality of those models. True errors, such as erroneous reactions, can produce misleading results. Incompleteness increases the need for completion and refinement. Incorrect syntax makes it more difficult to re-use the initial models with existing software tools. In the end, all of these issues translate into greater workload and time loss for the user. However, the quality of the models produced by the workflow crucially depends on the accuracy and completeness of the sources of information. If the pathway data are incorrect, there is little that an automatic conversion system can do beyond checking for feasible stoichiometries, mass and charge conservation and the like. Similarly, if some biological information is missing, the pathway-to-model workflow cannot easily create it. An example of this is information about compartmentalization. If the localization of the pathway nodes is not specified in the initial data, the resulting models will have a single compartment containing all molecular species.

**Figure 7** presents the size of the models produced by the project, in terms of number of state variables and number of mathematical relationships (i.e., reactions and transitions). The whole genome reconstructions present similar distributions for variables and relationships (**Figure 7A**). The situation is similar to the curated branch of BioModels Database (**Figure 7D**), which features models capable of numerical simulation. In contrast, the individual metabolic pathways (**Figure 7C**) are severely underdetermined, with many more variables than relationships. A possible reason for this is that entities in KEGG pathways are inferred by gene/enzyme homology, which can lead to missing reactions and therefore disconnected graphs.

## Systematic generation of genome-scale metabolic reconstructions from existing data resources

While the generation of genome-scale metabolic reconstructions typically relies upon time-consuming and manual efforts, techniques are being introduced which attempt to automate at least part of the process. One such approach to semi-automated reconstruction of such networks is that of the Model SEED [42]. This method provides a web-based resource for the generation of genome-scale metabolic reconstructions from assembled genome sequences. It has resulted in the generation of 130 (reported) reconstructions of a range of bacterial species, and has the potential for generating many more. While an approach that allows for the automated generation of reconstructions directly from the genome will clearly grow in importance given the ever-increasing volume of sequencing data, it is also clear that existing, curated data resources such as MetaCyc and KEGG still provide a great deal of biochemical knowledge that can be exploited in the metabolic reconstruction process. Many reconstruction projects take existing pathway databases such as these as a starting point, and indeed, recently introduced software tools such as the RAVEN Toolbox [48] have followed the examples set by the SuBliMinaL

Toolbox [39] and KEGGtranslator [49] in automating the generation of models from KEGG.

This work describes the first example in which an automated model reconstruction tool has been systematically applied to a wide range of organisms on such a scale. The result of this is the largest collection of genome-scale metabolic reconstructions to date. Due to their common formatting, use of identifiers and semantic annotations, the collection provides both a useful starting point for subsequent manual and semi-automated curation, and, as can be seen in the phylogenetic tree of **Figure 5**, a framework upon which metabolism can be systematically compared across species.

**Complementing pathway models with kinetic information**

Some aspects of the procedure described here compare with the work of Li and colleagues [50]. For instance, both their workflow and ours extract kinetic data from SABIO-RK. However, the aim of Li *et al.* was to provide full models, including parameterization and initial conditions. Their workflow could therefore plug in downstream of Path2Models' workflow; starting from models containing tentative rate-laws rather than stoichiometric reactions alone.

Even for the most extensively investigated organism, *Homo sapiens*, kinetic data is only available for 12.2% of its known metabolic reactions. Much less information is available for other organisms. It should be noted that despite the wealth of pathways and reactions gathered in databases such as KEGG or MetaCyc, they could still not claim to be comprehensive. The model presented here can therefore only reflect the knowledge available today in a re-usable form. Since kinetic equations (and parameters) have not been experimentally determined, there is a great interest in the application of generic approaches [51]. The modular rate laws suggested by Liebermeister *et al*. [52] have been specifically derived for cases in which more precise information remains elusive.

Each modular rate law can be used in three different modes or versions, which increase in complexity from the explicit (*cat*), through the Haldane-compliant (*hal*), to the Wegscheider-compliant (*weg*) version. These versions determine the form of the numerator in the equation (see Methods). A parsimonious approach was chosen in this work, where only as much complexity as necessary was introduced. Therefore, the most simple *cat* version of these rate laws was selected for all reversible reactions, even if this equation might not guarantee thermodynamic correctness. If the models created by this approach are used as the basis for subsequent calibration by experimental data, use of the *cat* version has two important advantages: (i) it contains a small number of parameters with uncertain values; and (ii) it has a low complexity in comparison to the *hal* or the *weg* version, with consequences on runtime. It should be noted that Liebermeister *et al*. have suggested an algorithm for transforming the parameter values of complex versions of the modular rate laws to the nearest simple form. It is possible to compute thermodynamically correct *cat*-parameters based on randomly selected *weg*-parameters through an intermediate step involving *hal*-parameters. However, application of this method would also require that all rate laws are re-created before and after parameter estimation.

Since the modular rate laws can only be applied to reversible metabolic reactions, it was therefore necessary to select further generic rate equations for the large-scale approach described in this work. It can be hoped that the percentage of experimentally determined rate laws will increase in the future , but generic rate laws will still be required to complete the quantitative models.

**Scaffold of logic models from KEGG signaling pathways**

As mentioned above, the automatically generated models are only partially parameterized.

In the case of KEGG signaling pathways for which no mechanistic details are provided, the models (with *qual* constructs) contain only topological relationships together with interaction signs. No logical rules specify the effects of (combined) interactions, and these models should be seen as scaffolds to be further parameterized before use in simulation. This can be done either by considering default, yet biologically meaningful, logical functions (e.g., requiring the presence of at least one activator and absence of all inhibitors) [53], by doing further manual refinement of the model (e.g., by literature mining), or by using dedicated experimental data to identify the functions [54].

Several simulation tools now support the SBML Level 3 *qual* package, including GINsim [55], CellNOpt [56] and the Cell Collective platform [57]. CellNOpt provides a pipeline to generate logical rules by pruning a general scaffold with all possible rules so as to find the submodel that best describes the data. This can be done using various formalisms [58] of increasing detail, depending of the data at hand. The Cell Collective platform includes Bio-Logic Builder to facilitate the conversion of biological knowledge into a computational model [59]. GINsim provides complementary features that allow performing multiple analyses of logical models using powerful algorithms [60]. Therefore, relying on a combined use of these tools, one could use the Path2Models qualitative models by training them against data of, for instance, a cell type of interest, and subsequently analyzing the resulting models.

**Creation of SBGN maps applying constraint-based layout**

SBGN provides a uniform and unambiguous graphical representation of biological knowledge. Providing models represented using this standard graphical format therefore facilitate visual human understanding. Some tools provide translation of SBML files into SBGN maps. However, to improve readability of such maps an appropriate layout of its elements is necessary. Here the initial positions of the model elements, extracted from the KEGG database graphical pathway representations, were used to produce layout of the SBGN maps. Although many general layout algorithms have been proposed in the last three decades [61,62], almost none of them support additional constraints such as predefined positions and spatial relationships that would be necessary to preserve the essence of the original KEGG maps. Therefore a constraint-based layout approach [63] in conjunction with orthogonal object-avoiding edge routing [64] was used. This allowed us to generate layouts without node overlaps and with improved readability while still preserving the overall structure of the map. Nevertheless, some open questions remain, such as the occasional presence of oversized labels in contrast to the uniform size of the glyphs, and long edges between glyphs. The impact of the latter issue could be reduced in subsequent versions by additional cloning of glyphs, involving the annotated multiplication of symbols representing the same entity, thus allowing this entity to be located at different points of the map.

**CONCLUSION**

All the software building blocks used in this project are freely available and can be used to build similar workflows. For instance, new modules can be used to read pathway information from other databases, as was shown for the entire PID [59]. As more sets of models are produced, they will be added to BioModels Database, where they will be easily retrievable and accessible. The availability of models in standard formats facilitates their import, comparison, merging and re-use. Automated development of models on the large scale will become crucial as automatic generation of pathways from genomics and metagenomics becomes common practise. Ready-made models will also be accurate starting points for the development of mechanistic models of whole cell models [60] where manual reconstruction is hardly an option. .

## METHODS

### KEGG pathways and the KEGG Markup Language

For the construction of quantitative kinetic models and qualitative models, the content of the KEGG PATHWAY database was obtained through its FTP site prior to 1 July 2011. Generic, reference pathways and organism-specific pathways for 1 515 specie were downloaded, all encoded in the KEGG Markup Language (KGML). These files mainly consist of *entries*, describing proteins and compounds of a pathway, and *interactions* between them. The *interactions* are subdivided into *reactions* and *relations*. *Reactions* correspond to biochemical reactions involving compounds and enzymes. *Relations* are used in the case of signaling pathways to specify protein-protein interactions. Layout information is given only for *entries* (i.e., nodes). Furthermore, each organism-specific pathway is derived from a reference pathway map. This involves adding organism-specific identifiers and setting the color (green) of enzymes that have protein instances in the current organism. Enzymes that have no known instance in an organism-specific pathway are retained in the map (albeit, while being colored differently) and keep their orthology identifier. This retention of absent enzymes is due to the focus of KGML files on visual representation of pathways rather than computational modeling. Completion and post-processing steps are therefore required to generate correct models from the KGML files [67].

Construction of the genome-scale metabolic reconstructions was performed through access of the publicly accessible KEGG web services, and was therefore applied to a more recent version of April 2013.

### Generation of SBML Level 3 *Core* from KEGG metabolic pathways

The generation of pathway models from KEGG information was performed with KEGGtranslator [49,67]. Each KGML *entry* was translated to an SBML Level 3 *species* (SBML *Core*) and an SBO term [68] was assigned (see **Table 1**). Each KGML *reaction* was translated to an SBML *reaction* (SBML *Core*). In addition to all substrates, products and catalyzing enzymes, this includes information about the reversibility of the reaction and the stoichiometry of each participant. Each reaction was checked against the KEGG API's reaction definition and missing reaction components and reaction modifiers (i.e., enzymes) were added to the model. The layout of each node (position, width and height) was also stored in the model, using the SBML *Layout* extension [69]. During the translation, enzymes that are contained in the orthologous template pathway, but have no instance in the current organism were removed from the model. Furthermore, for the metabolic translations, all nodes that do not correspond to physical instances of compounds or gene products were removed (i.e., pathway-reference nodes).

The models were augmented with Identifiers.org URI [70] cross-references to the following resources: 3DMET, ChEBI, DrugBank, Enzyme Nomenclature (EC code), Ensembl, Gene Ontology, GlycomeDB, HGNC, KEGG (gene, glycan, reaction, compound, drug, pathway, orthology), LipidBank, NCBI Gene, OMIM, PDBeChem, PubChem, Taxonomy, UniProt. Furthermore, every species, qualitative species, reaction and transition was assigned the ECO-code *ECO:0000313* meaning "a type of imported information that is used in an automatic assertion". If multiple identifiers from the same database could be assigned to a single element, BioModels.net biology qualifier [71] *has version* was used. Otherwise, BioModels.net biology qualifier *is* was used.

Additional information was stored in SBML *notes*, including a human-readable description (i.e., the full name), synonyms (different gene symbols, compound labels, etc.), pathways, and for small molecules, links to images of chemical compounds (hosted by KEGG and ChEBI), Chemical Abstract Service (CAS) numbers, chemical formula and molecular weight.

KEGG *groups* (which mostly correspond to complexes or gene families) were translated to species with all contained elements specified in the SBML *notes* and *annotation*. A human-readable list of contained gene symbols was added to the *notes*. A machine-readable term from a controlled vocabulary with a BioModels.net biology qualifier *is encoded by* was used to denote all group members.

**Generation of kinetics models for the metabolic networks**

The program SBMLsqueezer [72,73] was used to fetch kinetic equations from SABIO-RK. For all cases when a corresponding entry for a reaction in the model could be found in SABIO-RK, the rate law and kinetic parameters (including SBML values and *UnitDefinition* objects) were extracted. Corresponding entries within the SABIO-RK database were identified using the MIRIAM-compliant annotations of reactions within each model. SABIO-RK returns an SBML document that may contain several rate equations for the same reaction, depending on experimental conditions. For every rate law found in SABIO-RK, a correspondence was established between its species and compartments and those involved in the reaction of the query model. Functions and units defined by SABIO-RK that are referenced within the rate law of interest were also added to the model. In some cases such a matching was not possible. In these situations, the algorithm tries to add another rate law from SABIO-RK that matches the search criteria to the current reaction. The algorithm retains the order of rate laws as given by the search results from SABIO-RK. For the remaining reactions, either SABIO-RK could not find a rate equation or it was not possible to match species and compartments returned by SABIO-RK to the ones in the query model.

All missing rate laws were generated with the program SBMLsqueezer. To create *ab initio* kinetic laws for reversible enzyme-catalyzed reactions, the Common Modular (CM) rate law of Liebermeister *et al*. [52] was used. The explicit *cat* form was selected because it requires fewer independent parameters than the Haldane- (*hal* [74]) and Wegscheider-compliant (*weg* [75]) CM forms, described in more detail below. The CM rate law can be used for any kind of reversible enzyme-catalyzed metabolic reaction whose precise mechanism remains unknown. This is the case if rate laws are automatically created for all reactions in KEGG. In their work on the CM rate law, Liebermeister *et al.* also proposed four additional modular rate laws that all cover certain special cases.

A common denominator characterizes all modular rate laws. The precise structure of the denominator term depends on the number and type of involved modulators, such as inhibitors or stimulators, as well as the number of reactants and products. Each modular rate laws can be used in three different modes or versions: the explicit (*cat*), Haldane-compliant, and Wegscheider-compliant. These versions determine the form of the numerator in the equation. The *cat* version has the smallest number of parameters. Its numerator resembles the mass action rate law, but with each reacting species divided by its corresponding Michaelis constant. Equation (1) displays the *cat* version of the CM rate law with modulation function *f* that includes activations, inhibitions and effects of catalysts:

$$v_r(R_r, P_r, M_r, \vec{k}_r) = f(R_r, P_r, M_r, \vec{k}_r) \frac{k_r^+ \prod_{i \in R_r} \left(\frac{[S_i]}{K_{ri}}\right)^{h_r n_{ir}} - k_r^- \prod_{i \in P_r} \left(\frac{[S_i]}{K_{ri}}\right)^{h_r n_{ir}}}{\prod_{i \in R_r} \left(1 + \frac{[S_i]}{K_{ri}}\right)^{h_r n_{ir}} + \prod_{i \in P_r} \left(\frac{[S_i]}{K_{ri}}\right)^{h_r n_{ir}} - 1} \quad (1)$$

$R_r$, $P_r$, and $M_r$ denote the index sets for reactants, products and modifiers in the $r^{th}$ reaction, $n_{ir}$ gives the stoichiometric coefficient for the $i^{th}$ reactant, and vector $k$ contains all parameters, such as the Michaelis constant $K_{ri}$ and the cooperativity factors $h_r$. Multiplying the rate law with a well-defined prefactor function f allows the influence of modifiers, such as non-competitive inhibition to be included.

As mentioned above, modular rate laws are only defined for reversible enzyme-catalyzed reactions. **Table 2** summarizes the selected rate laws for irreversible reactions. In simple cases, the well-described Henri-Michaelis-Menten equation and the random-order ternary-complex mechanism were selected as the default rate law [76]. For arbitrary irreversible enzyme-catalyzed reactions, convenience rate laws [77] were created. These used the simpler thermodynamically dependent form when the stoichiometric matrix of the reaction system has full column rank, and the more complex thermodynamically independent form otherwise. For non-enzymatic reactions, the generalized mass action rate law [78] has been used. Effects of inhibitors or activators using the prefactor terms suggested by Liebermeister and Klipp were included. Just like the convenience rate law this equation can also be applied for arbitrary numbers of reactants and products and is therefore well suited for the automatic creation of unknown kinetic equations.

In order to keep the kinetic equations simple, a list of ions and small molecules to ignore when creating kinetic equations was defined. This is necessary to reduce the complexity of rate laws where their contribution would actually be limited (**Table 3**).

For gene-regulatory processes, the generalized version of Hill's equation [79] was selected. For species that are annotated as genes (SBO term identifier is a derivative of *gene*; SBO:0000), the *boundaryCondition* in the SBML definition of the *species* was set to *true*. This means that the concentration of genes is seen as a constant pool that cannot be influenced by reactions. Finally, in case of zeroth order reactions (i.e., reactions without any reactant or reversible reactions without any product), zeroth order versions of the generalized mass-action rate law were used.

The values of all new parameters were set to 1.0. The compartment sizes and species amounts or concentrations were also initialized with 1.0. If no substance, time, and volume units were defined in previous steps, the default substance unit was set to mole, time unit to second, and volume unit to litre. The units of all newly generated parameter objects were derived in order to ensure consistency of the overall models. This means that upon derivation, the units of reaction rates are all specified in substance per time. To this end, the SBML *hasOnlySubstanceUnits* attribute was set to *true* if it was undefined before, and species quantities that were given in concentration units were multiplied by the size of their containing compartment (within the kinetic equation) in order to obtain substance units for all species, irrespective if these were initially defined in concentration or substance units.

In order to facilitate the interpretation of the equations, units, and parameter objects created by this procedure, all elements were annotated with appropriate terms from SBO and the Unit Ontology [80].

**Development and implementation of SBML Level 3 *Qual* package**

Level 3 of SBML introduced the concept of modularity, with a *Core* package, shared by all, and domain-specific packages that add representational features on top of the core. The *qual* package is designed to provide SBML with the ability to encode qualitative models, such as logical models, or qualitative Petri-net models. The variables and the transformations of the models encoded in *qual* differ from species and reactions as defined in SBML *Core*. Qualitative models typically represent discrete levels of activities that are involved in transformations that cannot always be described as processes (consuming from and producing to pools of elements). To represent those concepts, *QualitativeSpecies* and *Transition* elements have been defined, together with their attributes and sub-elements. Briefly, a *QualitativeSpecies* encodes a variable representing a quantity or activity associated with an entity (e.g., gene, protein, but also phenomenological entity such as external condition, cell size, etc.) that can take discrete values (Boolean or multi-valued, e.g., in {0,1,2}). A *Transition* element encodes the rules governing the evolution of its *Output* node depending on the state of its *Input* nodes, both *Input* and *Output* nodes each referencing a particular *QualitativeSpecies* whilst providing additional information relating to the *Transition*. As most of the software packages used in this project were written in Java, JSBML [81] was chosen to implement the first library support for the SBML *qual* package. JSBML is a community-driven project to create a pure Java application programming interface (API) for reading, writing, and manipulating SBML files. It is an alternative to the Java interface provided in the C++ version, libSBML [82].

**Generation of SBML Level 3 *Qual* from KEGG signaling pathways**

The overall generation of SBML qualitative maps from KGML files was performed with KEGGtranslator [49,67] using an approach similar as used for kinetic models. Each KGML *entry* was translated to an SBML Level 3 *Qualitative Species* (*qual* package) and each KGML *relation* was translated in an SBML *Transition* (*qual* package).

In KGML, all interactions between two or more entities that are not molecular reactions are named KEGG *relations*. These relations describe enzyme-enzyme relations, protein-protein interactions, interactions of transcription factors and genes, protein-compound interactions and links to other pathways. The KEGG specification defines 16 different subtypes to describe the nature of the relations in more detail [83]. SBML *qual* describes relations as *Transitions*. *Transitions* consist of *Input*, *Output*, and *Term* objects. In contrast to KGML, SBML *qual* specifies the kind of relation in the attribute *sign* of the *Input*, instead of using type and subtype attributes for the relation. The *sign* attribute can take the values *positive* when the *qualitativeSpecies* linked to the input stimulates the transition, *negative* when it inhibits the transition, *dual* when the effects can go in both directions (depending upon the context), and *unknown*.

Before converting the KEGG pathway to SBML *qual*, the pathway relations were further enriched with BioCarta information distributed by the Nature Pathway Interaction Database [3], which provides human pathways in BioPAX Level 3 format. To this end, for each KEGG relation, a search for a corresponding BioCarta interaction was performed. Then, the relation was assigned to a new subtype depending on the BioCarta-ControlType attribute that can be activating or inhibiting.

For the conversion from KGML to SBML *qual*, the subtypes *activation* and *expression* are translated to the value *positive*. The subtypes *inhibition* and *repression* are translated to the value *negative*. All other subtypes are translated to the value *unknown*.

The value *dual* is assigned if a KEGG relation has both an activating as well as an inhibiting subtype. In addition to the sign attribute, the *Input* object is assigned an SBO term that further specifies the semantics based on subtype translated (see **Table 4**).

**Genome-scale metabolic reconstructions**

The genome-scale metabolic reconstructions were generated by applying a software pipeline based on modules of the SuBliMinaL Toolbox [39] and libAnnotationSBML [38] to all organisms in KEGG, release 66 (April 2013), accessed via the resource's web services interface. Many models were augmented with metabolic pathway information extracted from MetaCyc (version 17.0, March 2013), extending a previous approach that was applied to *Arabidopsis thaliana* [84]. In the cases of both KEGG and MetaCyc, this metabolic pathway information included metabolites, metabolic reactions and catalytic enzymes. Metabolites and reactions were reconciled with MNXref [40], and enzymes were specified with UniProt identifiers where possible.

The models do not contain any definitions of intracellular compartments. However, extracellular and intracellular compartments are specified, and a minimal extracellular growth medium was applied to all models, along with necessary transport reactions that allow for its uptake. The medium contains: $\alpha$-D-Glucose, $\beta$-D-Glucose, ammonium, sodium, potassium, magnesium, calcium, sulphate, chlorate, phosphate, protons, water, carbon dioxide and oxygen. Furthermore, default transport reactions have been added to allow for the transport of all intracellular metabolites into the extracellular space.

Commonly used biomass components were applied to each model, containing the 20 most common amino acids, the nucleotide precursors of RNA and DNA, glycogen and ATP, along with a default biomass reaction consisting of all 30 of these components. No attempt to tailor the biomass components to the organism was performed, and as such, clear anomalies such as the inclusion of glycogen in bacteria and plants remain. However, the removal of such terms, and the amendment of the biomass function itself, is a simple task for manual curation. All models were analyzed with the COBRA Toolbox [43] to determine whether they were able to synthesize the biomass components, with the results provided in **Supplementary Table 1**.

All source code and the compiled software application for generating genome-scale models is available in **Supplementary File 1**.

**The Systems Biology Graphical Notation**

The Systems Biology Graphical Notation [10] is a set of standard graphical languages for representing biological processes and interactions. The *Process Description* (PD) language allows scientists to represent chemical kinetics models, with pools of molecular entities consumed and produced by reactions. The *Activity Flow* (AF) language allows scientists to represent influence diagrams, in which entity activities inhibit or stimulate other entity activities.

**Generation of SBGN PD maps from SBML Level 3 *Core***

The generation of SBGN *Process Description* (PD) maps from SBML Level 3 *Core* and their subsequent automatic layout was performed with SBGN-ED [86]. Each SBML entry was translated to the corresponding SBGN PD glyph based on SBO terms (see **Table 2**). The original positions of the KGML elements, which were stored using the SBML *Layout* package, were used as initial positions for the SBGN PD glyphs. For each reaction, arcs to the corresponding reaction glyph connected the reaction partners. The types of the arcs, reflecting consumption, production or catalysis, were also set using SBO terms. Simple chemicals without a previously stored position or with more than one connection, along with all macromolecules with more than one connection, were cloned so that they appeared multiple times in the diagram, each with a connection to just a single element. The results of these steps were SBGN PD maps with valid structure but incomplete layout. The final layout of the maps was computed as a subsequent step.

For process glyphs representing reactions not contained in the original KEGG pathway, initial positions were calculated based on availability of reaction partners with layout information from KEGG: if these reaction partners were not available, the reactions were placed at the top of the map, otherwise the reactions were placed near to reaction partners with layout information. For macromolecules representing enzymes, initial positions were computed taking into account the positions of corresponding substrates, products and reaction glyphs. For simple chemicals representing secondary compounds, initial positions were computed such that these elements were grouped into substrates and products and placed close to the process glyph that represents the reaction. The automatic re-layout of the maps was done using a constrained-based approach [63] with orthogonal edge routing [64] for connections. Based on layout information stored in the model, geometric constraints were defined to preserve horizontal and vertical alignments, containment, as well as relative order of glyphs. Orthogonal object-avoiding edge routing was performed for all edges except the ones connecting glyphs representing secondary compounds and the corresponding process glyphs. The resulting edge routes are similar to those in the KEGG images available online. Edge nudging (moving apart overlapping parallel edges) was then applied to ensure that the edge routes conform to the SBGN layout rules.

The results of these steps were SBGN PD maps with a compact SBGN-conforming layout similar to the original KEGG layout. Finally, the maps were exported as SBGN-ML [87] and PNG image files, and stored in the BioModels Database.

**Generation of SBGN AF maps from SBML *Qual***

Analogous to SBGN *Process Description*, SBGN *Activity Flow* (AF) maps were generated by parsing glyph locations and size information from the original KEGG layout via the SBML *Layout* extension in the generated qualitative model files. Glyph and arc types were set on the basis of SBO terms. Glyphs having multiple positions in the original layout were added to the map only once at the best fitting position of the pre-defined set. Overlapping glyphs were spaced out using libvpsc [88] from the Adaptagrams project [89]. PNG renderings of the SBGN-ML files were created using PathVisio [90].

**Extension of BioModels Database to support the distribution of models**

In order to distribute the models produced by the project, several changes to the database software infrastructure were required. In order to manage models encoded in SBML Level 3 and using several SBML packages, the infrastructure has been upgraded to use the latest version of JSBML. The underlying pipeline (handling all models from their submission to their release) has been extended, and a new branch was created in order to accommodate the models. This separate branch was necessary because these automatically generated models are not expected to go through the normal curation and annotation phases, which are mainly manual processes. The schema of the database (which is used to store metadata about the models) had to be extended. The models themselves are stored in the file system. A custom structure has been devised in order to ensure acceptable access time (as too many files in a given folder puts a lot of stress on the file system). The resulting new branch is sufficiently generic to be able to store models coming from other similar projects. A generic system of categories was also created, in order to classify the models and provide a simple method for their browsing. This is currently used to handle the three main categories (metabolic, non-metabolic and whole genome metabolism) as well as the various sub-categories (such as *Photosynthesis* or *Caffeine metabolism* which have models for several organisms).

A model display facility was developed, providing access to information about the model, including the annotation of the *model* element and its associated notes. The model page offers the possibility to download the model (encoded in SBML) as well as its graphical representation (in PNG, SVG and SBGN-ML). A link to an online form provides a convenient way for users to report any issues they may encounter.

Finally, a tool was developed to automatically submit a large number of models. It is able to read the models, perform several checks and customize model files (mainly at the level of the *notes* and *annotations* of the *model* element) to ensure greater consistency, extract all the information necessary for their display, and store both metadata and models in the database and file system.

Several methods have been created for browsing the data. One can start from the list of all represented organisms, followed by individual pathways, such as *Photosynthesis* or [*Caffeine metabolism*](), and the display of a selected model. Alternatively, one can start with the three main categories of models (metabolic, non-metabolic, and whole genome metabolism), followed by the kind of models available in this category, then choose an organism and finally access the display of one model. In addition, a dedicated search engine is provided, allowing users to retrieve models based on textual queries. It relies on an index (generated using Lucene, http://lucene.apache.org/core/) of the content of all the models. A query expansion mechanism allows searches using Gene Ontology term names.

Three archives (one per main category) of all the models are available for downloading from the EBI's FTP servers.

## AVAILABILITY OF SUPPORTING DATA

All models generated by the project are available from BioModels Database [40].

Supplementary file 1 is provided through labarchives, DOI:10.6070/H4WH2MX0

Supplementary table 1 is through labarchives, DOI:10.6070/H4RR1W6P

## COMPETING INTERESTS

The authors declare they have no conflicts of interest.

## AUTHOR CONTRIBUTIONS

AD coordinated the work done by CW, FB, FM, RK, MR at ZBIT, contributed to JSBML including Layout and *qual* package, implemented an algorithm for unit derivation, and generated kinetic laws, parameters and units *ab initio*. AZ supervised the researchers at ZBIT. FB, FM and MvI contributed to the development of the SBML *qual* implementation. FB further contributed to the translation of signaling models to SBML and augmented them with additional information from BioCarta. CW contributed the source KGML models, implemented the metabolic and signaling conversions from KGML to SBML and generated the initial SBML models. He contributed to the implementation of multiple SBML extensions that are used within the scope of the manuscript. MR and RK implemented the SABIO-RK search. NS generated the genome-scale metabolic models, responded to reviewers' comments and edited the manuscript. PM and DBK assisted with the generation of genome-scale metabolic models. TC, MW and FS dealt with the representation of the models as SBGN PD maps and their automatic layout. MS generated SBGN AF-ML and graphical renderings of the SBML *qual* models. CC contributed to the discussions on SBML *qual* usage. NR contributed to the development of JSBML. CL, MG and NR contributed to BioModels Database. SK finalized the SBML *qual* specification. MH contributed to JSBML and the SBML *qual* specification. JSR helped to initiate the project, and contributed to the discussions on SBML *qual* usage. NS had the original idea to automate the creation of models from pathways. NLN initiated and coordinated the project and the manuscript. All authors contributed to the writing of the manuscript.

## ACKNOWLEDGEMENTS

NS and PM acknowledge support from the European Union FP7 project UNICELLSYS (grant number: 201142). MvI and PM received financial aid from the EU project BioPreDyn (ECFP7-KBBE-2011-5 Grant number 289434). MH and SK acknowledge support from the US National Institute of General Medical Sciences (grant number GM070923). CW, FB, FM, RK, AD, and AZ are grateful for financial support by the Federal Ministry of Education and Research (BMBF, Germany) in the projects Virtual Liver Network (grant number 0315756) and National Genome Research Network (NGFN-Plus, grant number 01GS08134). AD thanks the EU for funding his Marie Curie International Outgoing Fellowship within FP7 (project AMBiCon, 332020). PM acknowledges support from the US National Institute of General Medical Sciences (grant number GM080219) and the BBSRC (grant number BB/J019259/1). MvI, FB, FM, MS, NR received dedicated support from the EMBL-EBI. NS also thanks Ben Morris of the University of North Carolina at Chapel Hill for generously and freely making available his code for converting the NCBI Taxonomy flat files into a Newick tree, which was used to generate the phylogenetic tree of genome-scale models.

**FIGURE LEGENDS**

**Figure 1** Workflow leading from pathway descriptions to computational models. From the pathway databases on the left, information is extracted and encoded in SBML. Mathematical features, such as kinetic rate equations and flux bounds, are then added to each model, along with a graphical description. The completed models are all distributed through the BioModels Database. See Methods for a detailed explanation of each step.

**Figure 2** SBGN *Process Description* map of a pathway, cutout of the pathway and parts of the SBML file describing the reactions shown in the cutout.

**Figure 3** Rate equations from SABIO-RK for models from selected organisms.

**Figure 4** Workflow indicating the SuBliMinaL Toolbox modules that were linked to produce draft metabolic models from the source data. *KEGG extract* and *MetaCyc extract* produce MIRIAM-annotated SBML representations of the contents of KEGG and MetaCyc, respectively. Metabolite and reaction ids are reconciled through reference to the MNXref namespace, unifying the metabolites to an assumed intracellular pH of 7.3, and mass and charge balancing reactions where possible. The Merge module merges the individual reconstructions from KEGG and MetaCyc, to which a limited growth medium and transport reactions are added, along with gene-protein relationships (GPRs) and flux bounds. The models are then formatted to allow for their analysis with the COBRA Toolbox and then released as draft models that represent the union of the information held in both KEGG and MetaCyc.

**Figure 5** Phylogenetic tree illustrating all 2 630 genome-scale metabolic models. The tree is color coded, indicating the presence of archaea, bacteria and eukaryota in the collection. Analysis results of each model are displayed, with bars indicating the number of metabolic reactions, metabolites, makeable metabolites and makeable biomass components in blue, red, purple and green respectively. In this illustration, the bars have been scaled for ease of visualization.

**Figure 6** A zoomed in view of the eukaryotic branch of the phylogenetic tree of Figure 5. The online iTOL web application version of the tree, available at [40], allows for zooming, searching and visualization of the tree and its associated statistics.

**Figure 7** Distribution of the models generated by the project according to their size, in terms of the number of molecular species (blue) and the number of mathematical relationships – i.e. reactions, transitions, rules etc. (salmon) in each class. A-C: the whole genome reconstructions, qualitative models, and chemical kinetic models. D-E: the curated and non-curated literature-based branches of the BioModels Database.

**Figure 8** Relative sizes of the different classes of models, based on their main Gene Ontology (GO) annotations. The GO terms annotating the SBML *Model* element for each model generated by the project were collected, and clustered to generate groups of models covering (what are considered therefrom to be) the same domain of biology.

**Table 1** | KGML entry type and corresponding mapping to SBO term.

| KGML entry type | SBO identifier | SBO name |
|---|---|---|
| compound | SBO:0000247 | simple chemical |
| enzyme | SBO:0000252 | polypeptide chain |
| gene | SBO:0000252 | polypeptide chain |
| ortholog | SBO:0000252 | polypeptide chain |
| group | SBO:0000253 | non-covalent complex |
| map | SBO:0000552 | reference annotation |

**Table 2** | Rate-laws for irreversible reactions

| Type of irreversible reaction | Rate law |
|---|---|
| non-enzyme reaction | Generalized mass action rate law |
| uni-uni enzyme reaction | Henri-Michaelis-Menten equation |
| bi-uni enzyme reaction | Random-order ternary-complex mechanism |
| bi-bi enzyme reaction | Random-order ternary-complex mechanism |
| arbitrary enzyme reaction | Convenience rate law |

**Table 3** | Small molecules and ions with negligible impact on reaction velocities

| Name | Formula | KEGG identifier |
|---|---|---|
| Water | $H_2O$ | C00001 |
| Zinc cation | $Zn^{2+}$ | C00038 |
| Copper(II) | $Cu^{2+}$ | C00070 |
| Calcium cation | $Ca^{2+}$ | C00076 |
| Hydron | $H^+$ | C00080 |
| Cobalt ion(II) | $Co^{2+}$ | C00175 |
| Potassium cation | $K^+$ | C00238 |
| Hydrogen | $H_2$ | C00282 |
| Nickel | Ni | C00291 |
| Hydrochloric acid | HCl | C01327 |
| Hydrogen selenide | $H_2Se$ | C01528 |
| Iron(II) ion | $Fe^{2+}$ | C14818 |
| Iron(III) ion | $Fe^{3+}$ | C14819 |

**Table 4** | KGML subtypes and the corresponding SBML *Qual* sign attributes and SBO identifiers

| KGML subtype | SBML *Qual* sign | SBO identifier | SBO name |
| --- | --- | --- | --- |
| activation | positive | SBO:0000170 | stimulation |
| inhibition | negative | SBO:0000169 | Inhibition |
| expression | positive | SBO:0000170 | stimulation |
| repression | negative | SBO:0000169 | inhibition |
| indirect effect | unknown | SBO:0000344 | molecular interaction |
| state change | unknown | SBO:0000168 | control |
| binding/association | unknown | SBO:0000177 | non-covalent binding |
| dissociation | unknown | SBO:0000177 | non-covalent binding |
| missing interaction | unknown | SBO:0000396 | uncertain process |
| phosphorylation | unknown | SBO:0000216 | phosphorylation |
| dephosphorylation | unknown | SBO:0000330 | dephosphorylation |
| glycosylation | unknown | SBO:0000217 | glycosylation |
| ubiquitination | unknown | SBO:0000224 | ubiquination |
| methylation | unknown | SBO:0000214 | methylation |

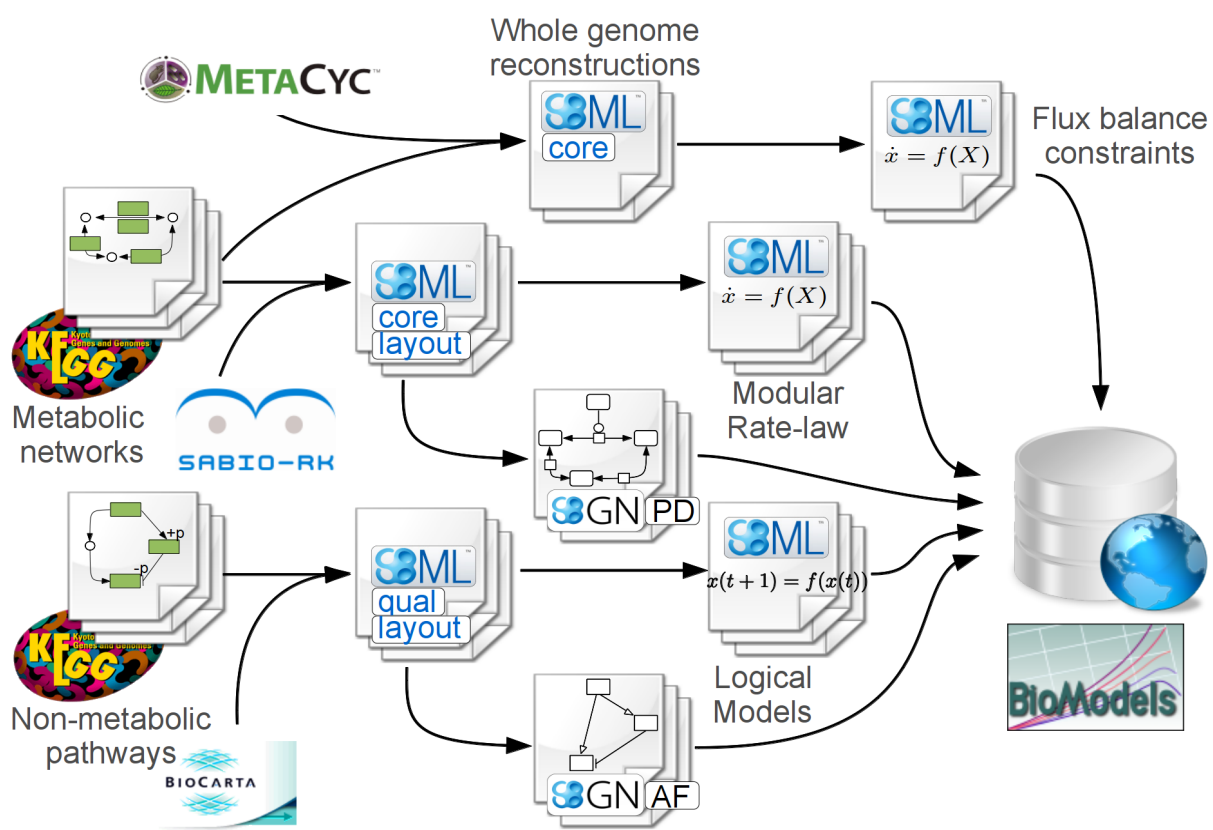

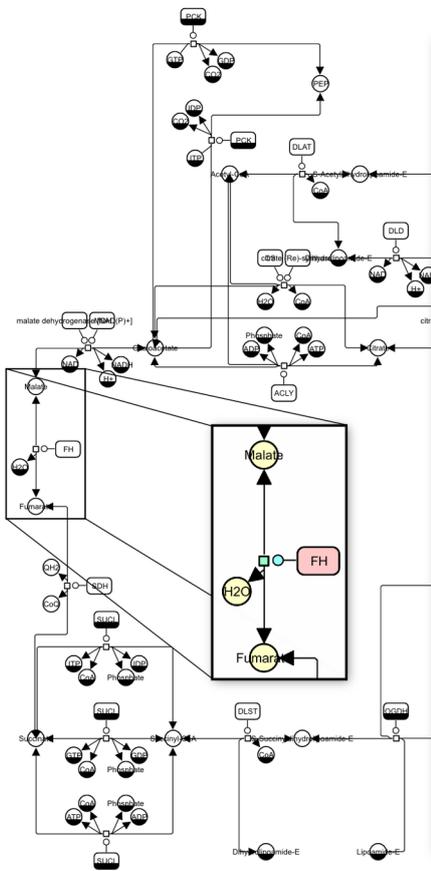

```xml
<?xml version='1.0' encoding='UTF-8' standalone='no'?>
<sbml xmlns="http://www.sbml.org/sbml/level3/version1/core"
      xmlns:layout="http://www.sbml.org/sbml/level3/version1/layout/version1"
      level="3" version="1" layout:required="false">
  <model id="BMID000000101169" name="Citrate cycle" metaid="meta_path_hsa00020"
         timeUnits="time" substanceUnits="substance" volumeUnits="volume">
    ...
    <listOfSpecies>
      ...
      <species id="FH" constant="false" initialAmount="1" hasOnlySubstanceUnits="false"
               name="FH" metaid="meta_FH" boundaryCondition="false" sboTerm="SBO:0000252">
        <annotation>
          ...
          <bqbiol:isEncodedBy>
            <rdf:Bag>
              <rdf:li rdf:resource="http://identifiers.org/ensembl/ENSG00000091483"/>
            </rdf:Bag>
          </bqbiol:isEncodedBy>
          ...
        </annotation>
      </species>
      ...
      <species id="Malate" constant="false" initialAmount="1" hasOnlySubstanceUnits="false"
               name="Malate" metaid="meta_Malate" boundaryCondition="false" sboTerm="SBO:0000247">
        ...
        <rdf:li rdf:resource="http://identifiers.org/obo.chebi/CHEBI:18012"/>
        ...
      </species>
      ...
    </listOfSpecies>
    ...
    <listOfReactions>
      ...
      <reaction reversible="true" sboTerm="SBO:0000176" fast="false" compartment="default">
        <annotation>
          ...
          <rdf:li rdf:resource="http://identifiers.org/kegg.reaction/R01082"/>
          ...
        </annotation>
        <listOfReactants>
          <speciesReference species="Malate" name="cpd:C00149" stoichiometry="1" sboTerm="SBO:0000010" />
        </listOfReactants>
        <listOfProducts>
          <speciesReference species="Fumarate" stoichiometry="1" sboTerm="SBO:0000010" />
          <speciesReference species="H2O" stoichiometry="1" sboTerm="SBO:0000010" />
        </listOfProducts>
        <listOfModifiers>
          <modifierSpeciesReference species="FH" sboTerm="SBO:0000040" />
        </listOfModifiers>
        <kineticLaw> .... </kineticLaw>
      </reaction>
      ...
    </listOfReactions>
  </model>
</sbml>
```

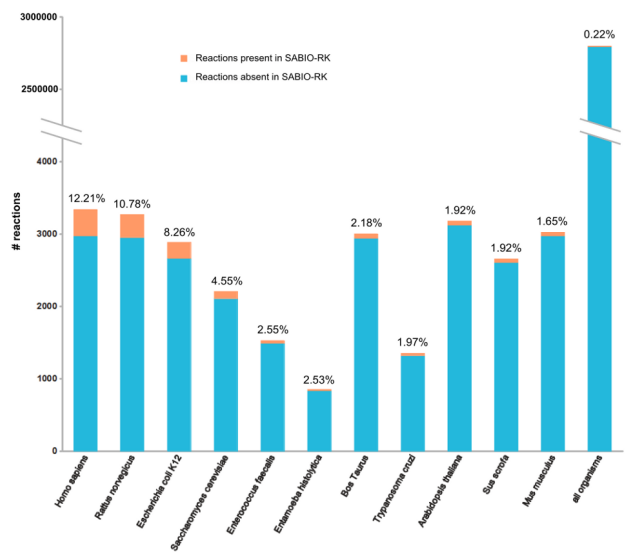

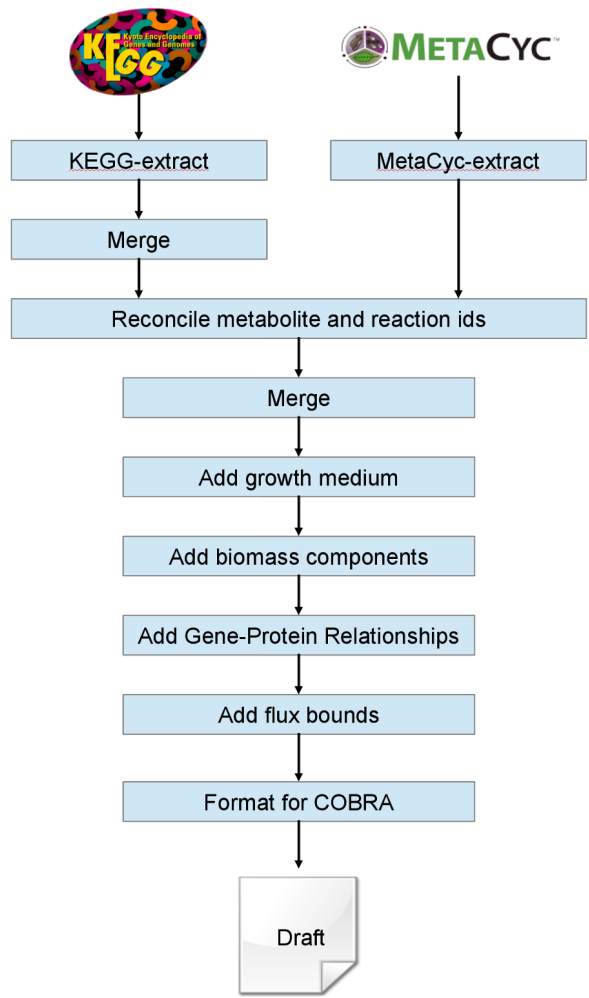

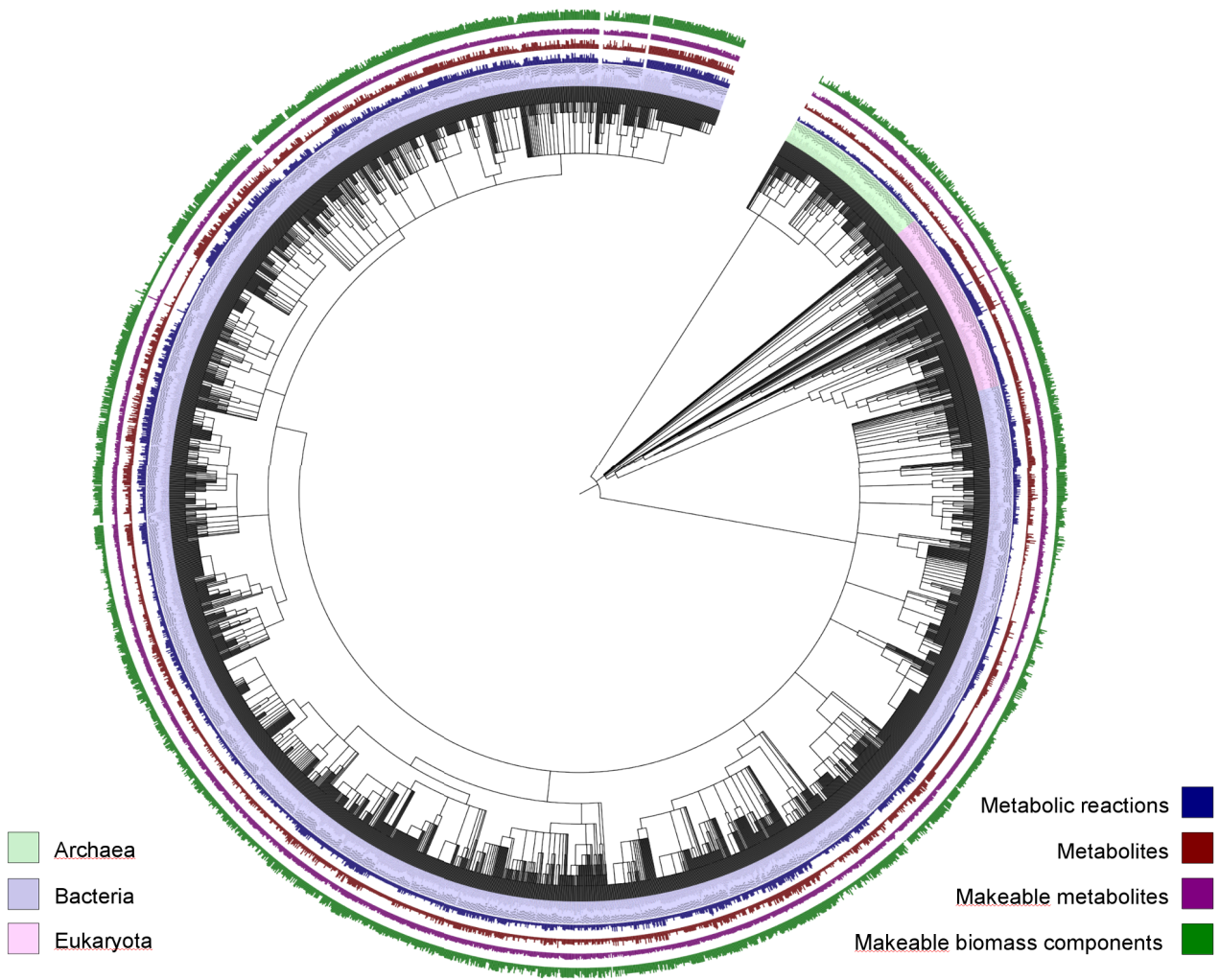

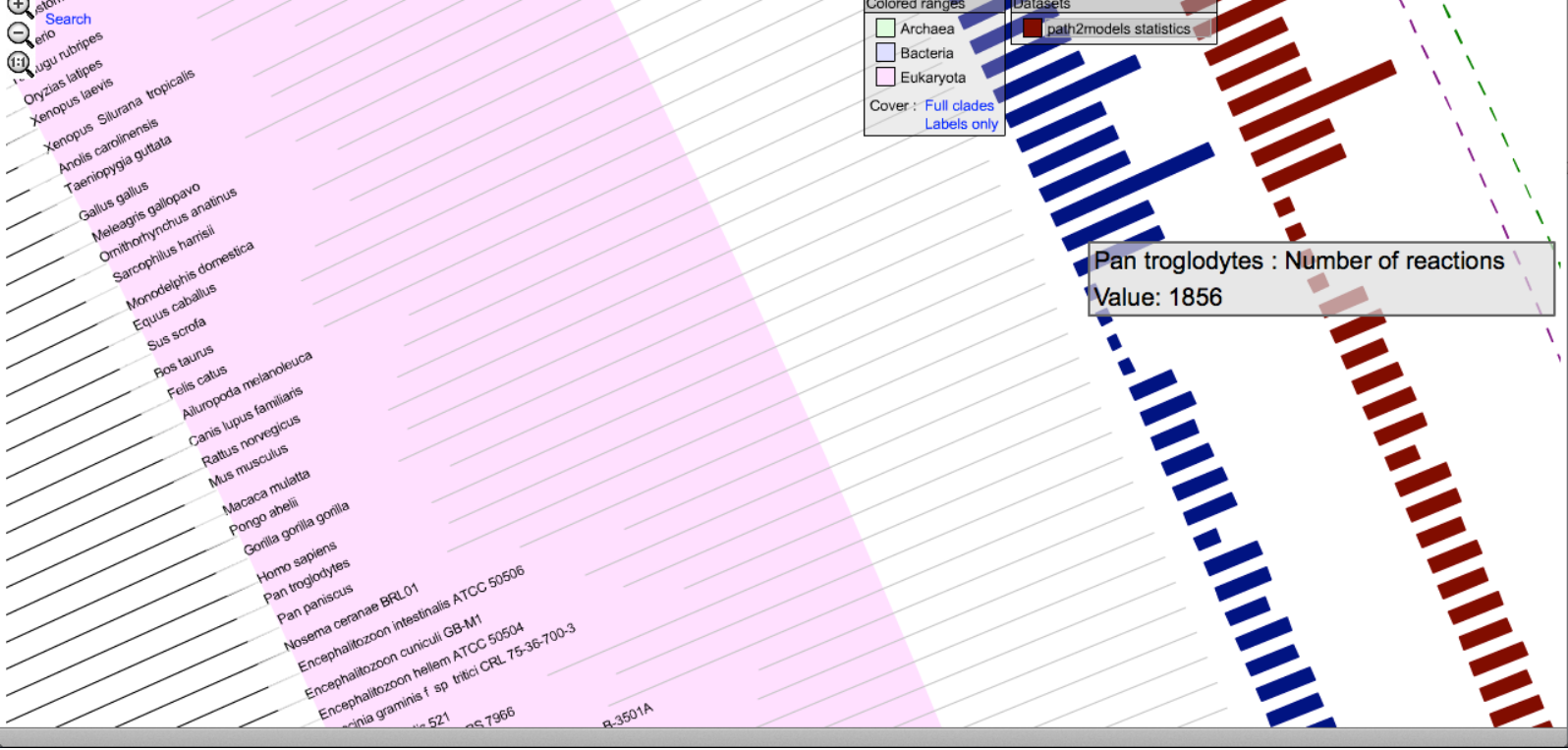

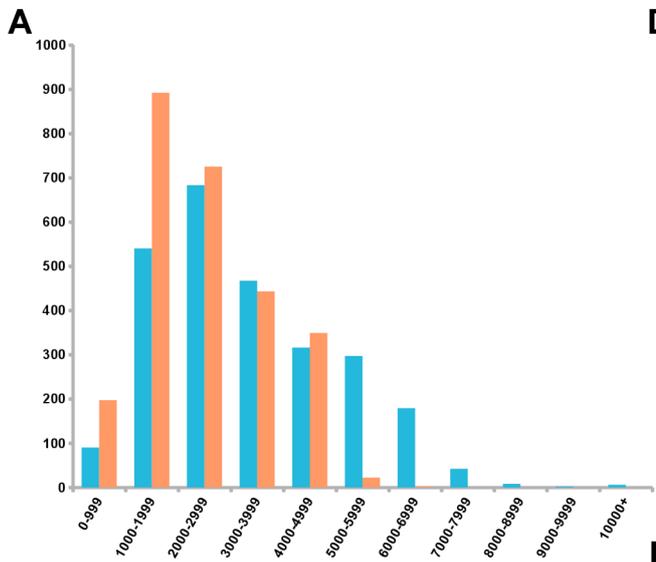
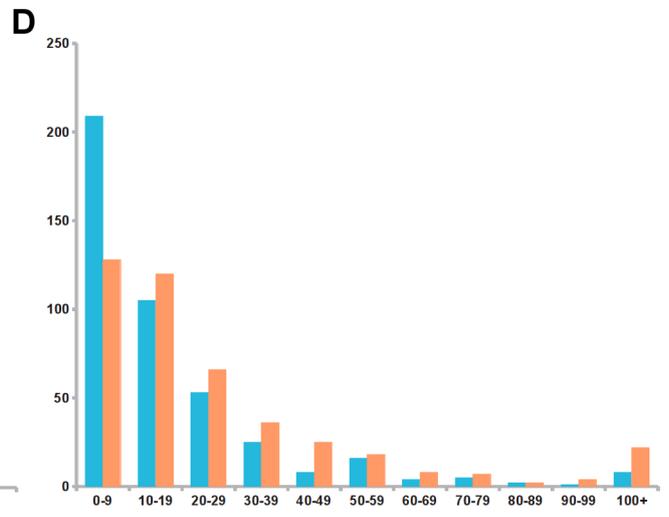
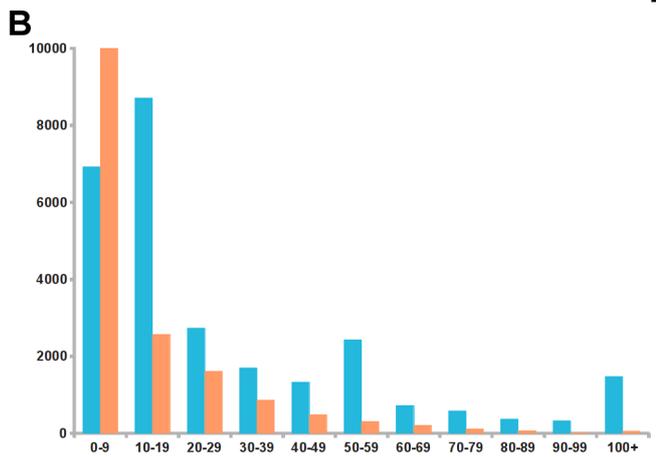
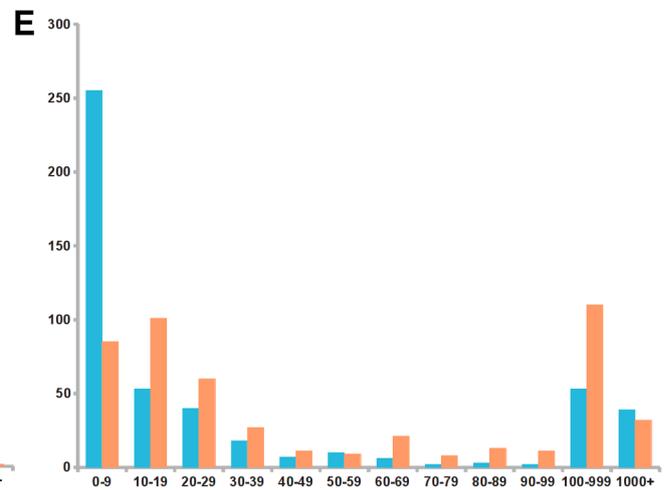
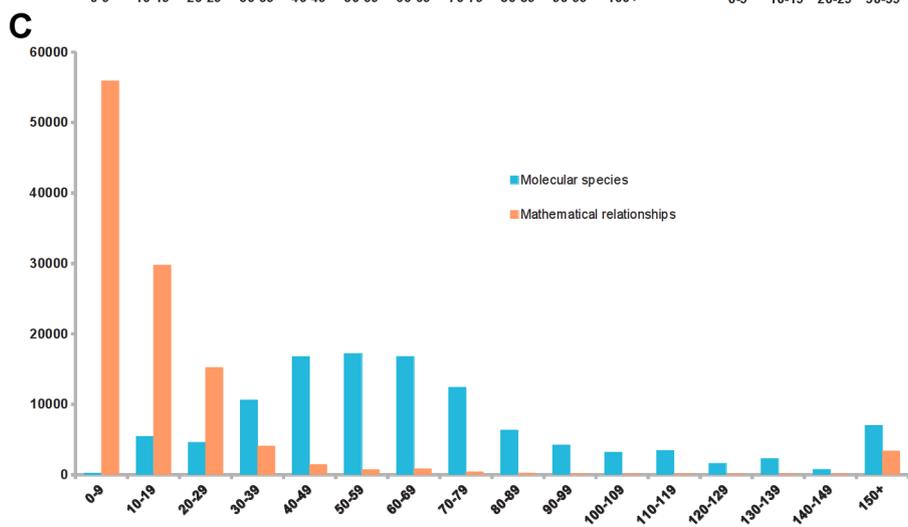

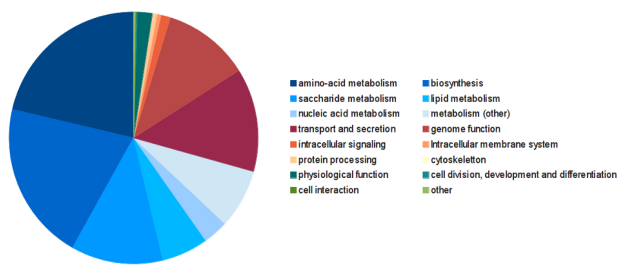